\documentclass[aps,prl,
reprint,
twocolumn,
10pt,
floatfix,
]{revtex4-1}
\usepackage{amsmath}    

\usepackage{graphicx}   
\usepackage{hyperref}
\newcommand{\jun}{junction }
\newcommand{\juns}{junctions }
\newcommand{\Jos}{Josephson }

\begin{document}
\title[R.Monaco \textit{et al.}]{The Critical Current of Point Symmetric Josephson Tunnel Junctions}
\author{Roberto Monaco}
\affiliation{CNR-ISASI, Institute of Applied Sciences and Intelligent Systems ''E. Caianello'', Comprensorio Olivetti, 80078 Pozzuoli, Italy}
\email[Corresponding author e-mail:$\,$]{$\,$r.monaco@isasi.cnr.it}
\date{\today}

\begin{abstract}
The physics of Josephson tunnel junctions drastically depends on their geometrical configurations. The shape of the junction determines the specific form of the magnetic-field dependence of the its Josephson current. Here we address the magnetic diffraction patterns of specially shaped
planar Josephson tunnel junctions in the presence of an in-plane magnetic field of arbitrary orientations. We focus on a wide ensemble of junctions whose shape is invariant under point reflection. We analyze the implications of this type of isometry and derive the threshold curves of junctions whose shape is the union or the relative complement of two point symmetric plane figures. 

\end{abstract}
\maketitle


Any \Jos device is characterized by a maximum zero-voltage d.c. current, $I_c$, called critical current, above which it switches to a finite voltage. How the critical current modulates with an external magnetic field is an important issue for all the earlier \cite{barone} and novel \cite{novel} applications of the \Jos effect. It has long been addressed that the magnetic diffraction pattern (MDP) of planar \Jos tunnel \juns (JTJs) drastically depends on both the specific shape of the tunneling area and the direction of the in-plane applied field. Most of the milestone works which allowed significant advances in the understanding of the geometrical properties of the MDP \cite{peterson,lemke,kikuchi} just considered a number of interesting shapes with the magnetic field applied in a preferential direction. However, the knowledge of the MDP for arbitrary field direction allows to evaluate the consequences of an unavoidable field misalignment in the experimental setups. Moreover, the measurements of $I_c(H)$ provides the first quality test of any Josephson device.   

In this Letter we highlight the MDP properties of a wide class of JTJs characterized by a point symmetric shape in the general case of an arbitrarily oriented in-plane magnetic field. Furthermore, in force of the additive property of the surface integrals, we show that the threshold curves of JTJs with complex shapes can be expressed as a linear combination of the MDP of junctions with simpler point symmetric shapes.
\vskip 8pt

In Josephson's original description the quantum mechanical phase difference, $\phi$, across the barrier of a generic two-dimensional planar \Jos tunnel \jun is related to the magnetic field, ${\bf H}$, inside the barrier through \cite{brian}:

\begin{equation}
\label{gra}{\bf \nabla} \phi =
\kappa{\bf H}\times {\bf u}_z ,
\end{equation}
\noindent in which ${\bf u}_z$ is a unit vector orthogonal to the \jun plane and $\kappa\equiv 2\pi\mu_0 d_m/\Phi_0$, where $\Phi_0$ is the magnetic flux quantum, $\mu_0$ the vacuum permeability, and $d_m$ the \jun \textit{magnetic} penetration depth \cite{wei,SUST13a}. The external field ${\bf H}$, in general, is given by the sum of an externally applied field and the self-field generated by the current flowing in the junction. If the junction dimensions are smaller than the \Jos penetration length, the self-magnetic field is negligible, as has been first shown by Owen and Scalapino \cite{owen} for a rectangular JTJ . Henceforth, for (electrically)  small JTJs the phase spatial dependence is obtained by integrating Eq.(\ref{gra}); in Cartesian coordinates, for an in-plane magnetic field applied at an arbitrary angle ${\theta}$ with the $Y$-axis, ${\bf {H}}=(H \sin\theta,H\cos\theta)$, it is: 

\begin{equation}
\phi(x,y,H,\theta,\phi_0)=\kappa H(x\cos\theta-y\sin\theta)+\phi_0,
\label{small}
\end{equation}
\noindent where $\phi_0$ is an integration constant. The tunnel current flows in the $Z$-direction and the local density of the \Jos current is \cite{brian}:

\begin{equation}
\label{jj}J_J(x,y,H,\theta,\phi_0) =J_c \sin \phi(x,y,H,\theta,\phi_0) ,
\end{equation}

\noindent where the maximum \Jos current density, $J_c$, is assumed to be uniform over the junction area. The \Jos current, $I_J$, through the barrier is obtained integrating Eq.(\ref{jj}) over the junction surface, $S$:
\begin{equation}
\label{IJ}
I_J(H,\theta,\phi_0)=\int_S J_J\,dS=J_c \int_S \sin \phi\,dS.
\end{equation}

\noindent  The \jun critical current, $I_c$, is defined as the largest possible \Jos current, namely,:

\begin{equation}
\label{Ic}
I_c(H,\theta)= \max_{\phi_0} I_J(H,\theta,\phi_0),
\end{equation}

The integral in Eq.(\ref{IJ}) applied to the surface of an axis-parallel rectangle yields:

\begin{equation}
\int_{y_1}^{y_2}\!\!\!\!\! dy\!\!\int_{x_1}^{x_2} \!\!\!\!\!\! dx \sin(k_x x- k_y y+ \phi_o)=\frac{4}{k_x k_y}\sin \frac{k_x(x_2-x_1)}{2} \times
\label{integral}
\end{equation}
$$\times\,\sin \frac{k_y(y_2-y_1)}{2} \sin \left[\frac{k_x(x_2+x_1)}{2}-\frac{k_y(y_2+y_1)}{2}+\phi_0\right],$$

\noindent where $(x_1,y_1)$ and $(x_2,y_2)$ are the coordinates of, respectively, the lower left and upper right rectangle corners.  Identifying $k_x$ and $k_y$ with, respectively, $\kappa H \cos\theta$ and $\kappa H \sin\theta$, the critical current is achieved when $2\phi_0=\pm \pi+k_y(y_2+y_1) -k_x(x_2+x_1)$ and we end out with the well-known double Fraunhofer diffraction pattern of a small rectangular JTJ in a arbitrarily oriented magnetic field \cite{barone}:

\begin{equation}
\label{IcRectangle}
I_c^R(H,\theta)\!\!=\!\! J_c A_R \!\left|\frac{\sin[(\kappa H W \sin\theta)/2]} {(\kappa H W \sin\theta)/2}\, \frac{\sin[(\kappa H L \cos\theta)/2]}{(\kappa H L \cos\theta)/2} \right|,
\end{equation}

\noindent where $W=x_2-x_1$ and $L=y_2-y_1$ are the rectangle edges and $A_R=WL$ its area \cite{note1}. The quantity within the absolute-value bars can be thought of as characteristic area-independent function, $\mathcal{F}_R(H, \theta)$, of all the rectangles with aspect ratio $W/L$:
\begin{equation}
\label{FR}
\mathcal{F}_R(H,\theta)\equiv \frac{\sin[(\kappa H W \sin\theta)/2]} {(\kappa H W \sin\theta)/2}\, \frac{\sin[(\kappa H L \cos\theta)/2]}{(\kappa H L \cos\theta)/2}.
\end{equation}


\vskip 8pt

To state the problem, let us consider a small planar JTJ whose tunneling area is obtained as the difference between two concentric and parallel rectangles of arbitrary aspect ratios such that the smaller rectangle, $r$, lies wholly inside the outer rectangle, $R$ ($r \subset R$). This geometry is depicted in Figure~\ref{rectannulus} where the origin of the Cartesian axes is in the center of the junction; the sides lengths of the outer (inner) rectangle are $W$($w$) and $L$($l$). Such a structure can be realized by etching away or anodizing the inner part of the junction base electrode. The surface of this complementary \jun can be decomposed in the four rectangles shown in gray; we can, therefore, calculate its MDP, $I_c^{R-r}(H,\theta)$, making use of Eq.(\ref{integral}) for a single rectangle and exploiting the additive property of integrals. After some algebraic manipulation, one founds that for any value of the field angle $\theta$:

\begin{figure}[t]
\centering
\includegraphics[width=8cm]{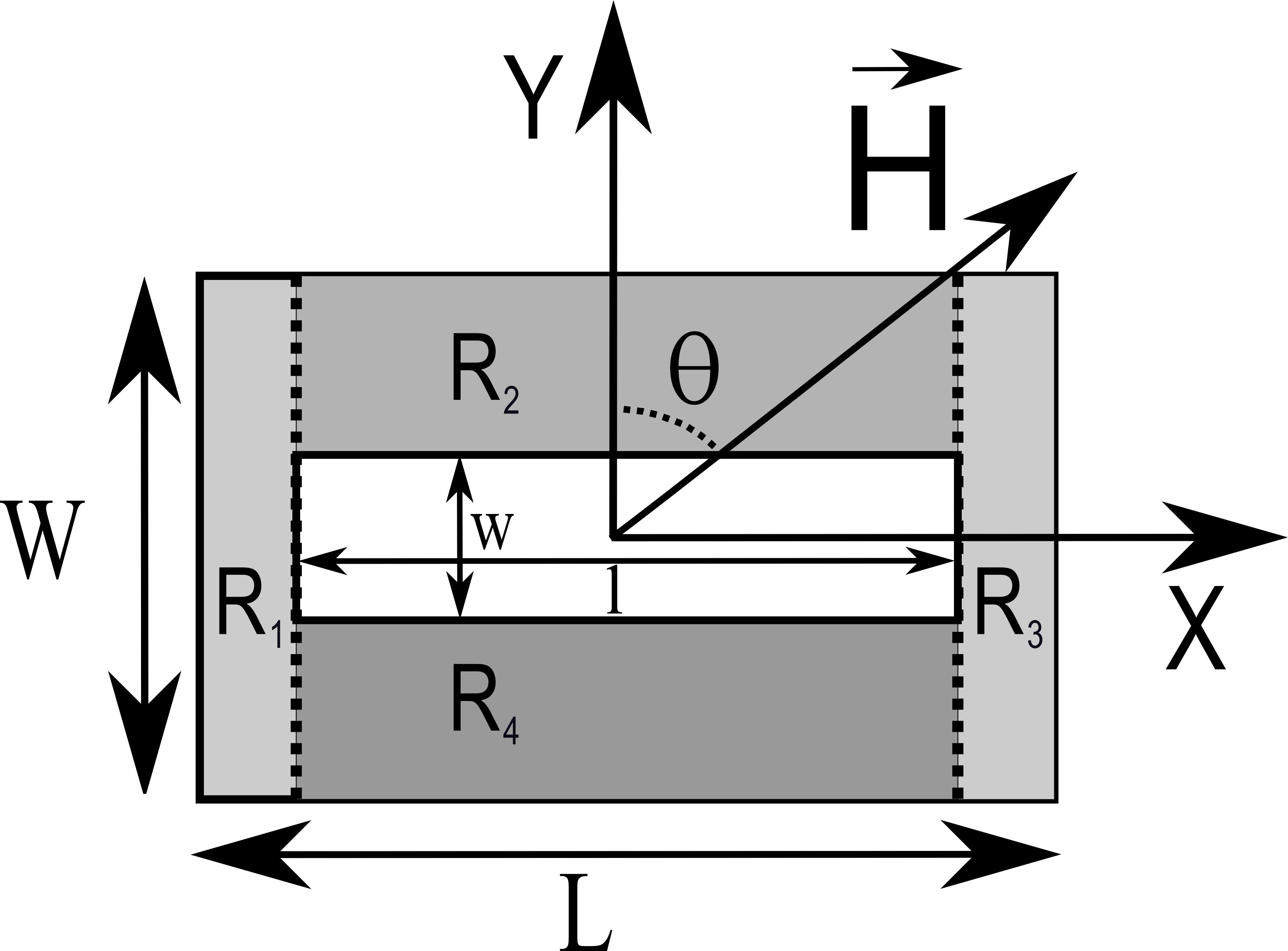}
\caption{Schematic of a complementary planar JTJ resulting by the difference between two concentric and axis-parallel rectangles of arbitrary aspect ratios. The outer rectangle has sides of lengths $L$ and $W$, while the inner one has sides of lengths $l$ and $w$. The junction area, $WL-wl$, is given by the sum of the areas of the four gray rectangles. The in-plane magnetic field, ${\bf {H}}$, is applied at a generic angle, $\theta$, with the $Y$-axis.}
\label{rectannulus}
\end{figure}

\begin{equation}
I_c^{R-r}(H,\theta)=J_c\left|A_R \mathcal{F}_R(H, \theta) - A_r \mathcal{F}_r(H, \theta) \right|,
\label{concRect}
\end{equation}

\noindent where $A_R=WL$, $A_r=wl$ and the characteristic functions $\mathcal{F}_{r}$ and $\mathcal{F}_{R}$ are defined through the MDPs of the fictitious inner and outer rectangular junctions, respectively, $I_c^{r}(H,\theta)\equiv J_c A_r \left|\mathcal{F}_{r}(H, \theta)\right|$ and $I_c^{R}(H,\theta)\equiv J_c A_R \left|\mathcal{F}_{R}(H, \theta)\right|$. Interestingly, Eq.(\ref{concRect}) still holds if the rectangles are not parallel \cite{note1}. Moreover, a similar expression also applies when one or both the rectangles are replaced by arbitrarily oriented concentric rhombuses or ellipses. 

\noindent We remind that i) for a small diamond-like JTJ of diagonals $P$ and $Q$ parallel to Cartesian axes the characteristic function is \cite{nappi96}:

\begin{equation}
\label{FD}
\mathcal{F}_D(H,\theta)=2 \frac{\cos[(\kappa H P \sin\theta)/2] - \cos[(\kappa H Q \cos\theta)/2]} { (\kappa H P \sin\theta/2)^2 - (\kappa H Q \cos\theta/2)^2} 
\end{equation}
\noindent and ii) for a planar JTJ delimited by an axis-aligned ellipse of principal semi-axes $a$ and $b$ it is \cite{ekin}:

\begin{equation}
\label{FE}
\mathcal{F}_E(H,\theta)= 2 \frac{J_{1}\!\left[\kappa H p_E(\theta)/2\right]}{\kappa H p_E(\theta)/2},
\end{equation}

\noindent where $J_1$ the $1$st order Bessel function of the first kind and $p_E(\theta) \equiv 2 \sqrt{ a^2 \sin^2 \theta+b^2 \cos^2 \theta}$ the length of the projection of the ellipse in the direction normal to the externally applied magnetic field. Eq.(\ref{FE}), first reported by Peterson \textit{et al.} \cite{ekin} in 1990, generalizes the so called \textit{Airy pattern} of a circular junction \cite{barone} of radius $r=a=b$. 

\vskip 2pt

Indeed, we found that the MDP of a complementary JTJ resulting from the difference (complement), $s'=S-s$, of two concentric plane figures with two lines of symmetry (including unconventional shapes like, for example, crosses, bow-ties, s-shapes and figure-eights), can be expressed in terms of their areas, $A_S$ and $A_s$, and characteristic functions, $\mathcal{F}_S$ and $\mathcal{F}_s$, that is: 

\begin{equation}
I_c^{s'}(H,\theta)=J_c\left|A_S \mathcal{F}_S(H, \theta) - A_s \mathcal{F}_s(H, \theta) \right|.
\label{diff}
\end{equation} 

\noindent A similar expression was proved for the sum (union), $S$, of two disjoint figures, $s$ and $s'$, namely: 

\begin{equation}
I_c^{S}(H,\theta)=J_c\left|A_s \mathcal{F}_s(H, \theta) + A_{s'} \mathcal{F}_{s'}(H, \theta) \right|.
\label{sum}
\end{equation} 

In the following, we will demonstrate that the broadest geometrical requirement for the validity of Eqs.(\ref{diff}) and (\ref{sum}) is the point-symmetry of the plane figures.

\vskip 8pt
%
%
\begin{figure}[t]
\centering
\includegraphics[width=8cm]{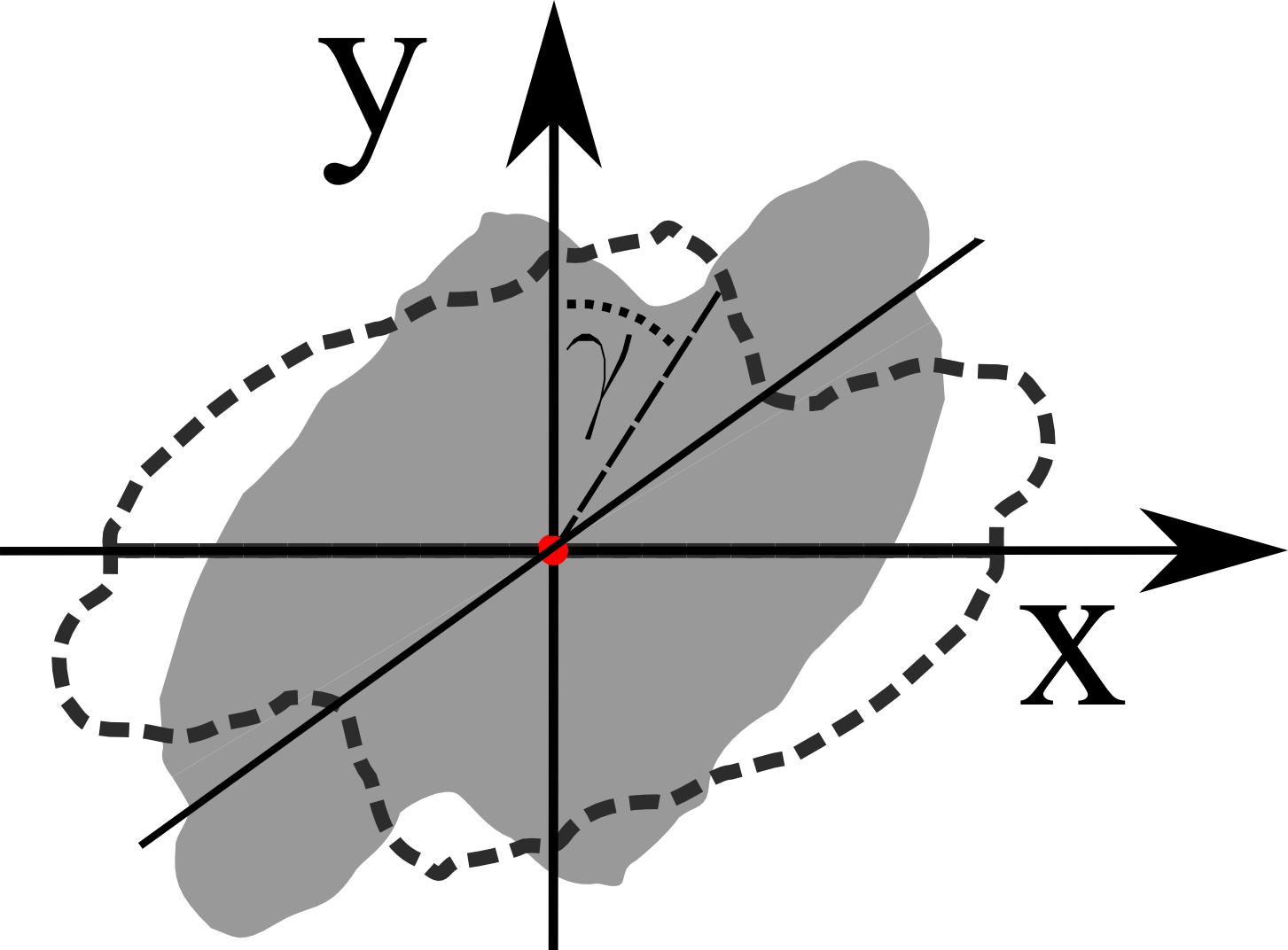}
\caption{Point-symmetric plane figure in gray; the center of symmetry is in the origin of the Cartesian system. The dashed line is the contour of the same figure when rotated by an arbitrary angle $\gamma$.}
\label{pointsymmetric}
\end{figure}

Let us consider a small JTJ whose shape has a second order \textit{point} or \textit{central-inversion} symmetry, that is to say, is invariant upon a $180^\circ$ rotation around one point called center of symmetry, namely upon reflections in two perpendicular lines. If we pick any Cartesian system with origin in the center of symmetry, then the figure contour in the first (second) quadrant is reproduced specularly in the third (fourth) quadrant. One example of point-symmetric figure is illustrated by the gray shape in Figure~\ref{pointsymmetric}. Any line drawn through the center of symmetry crosses the figure in two equidistant points. If we rotate the figure of any arbitrary angle, $\gamma$, around its center of symmetry, we still obtain a point symmetric figure (see the dashed contour in Figure~\ref{pointsymmetric}). Furthermore, if the $S$ and $s\subset S$ are two concentric point symmetric figures, the relative complement of $s$ in $S$ is also point symmetric. The category of point symmetric plane figures is wider that the set of figures with two perpendicular lines of symmetry; more generally, it includes regular digons (degenerate polygons with two edges and two vertices), parallelograms and other polygons having an even number of sides with opposite sides equal in length and parallel, to mention a few non-trivial ones.

\vskip 8pt

\noindent By way of example of the use of Eq.(\ref{sum}) applied to point symmetric JTJs, a \begin{large}$+$\end{large}-shape can be considered as the union of two point symmetric figure: a rectangle and the union of two disjoint squares. (Generalizing, an \begin{large}x\end{large}-shaped surface is a point symmetric figure regardless of the angle between its legs; it results from the union of a parallelogram with the union of two disjoint trapezoids.)

\vskip 8pt

For a point symmetric JTJ, in force of the sine function oddity, the surface integral over its surface $S$ of $\sin(k_x x -k_y y)$ is automatically zero; therefore, the $\phi_0$-dependence of the \Jos current in Eq.(\ref{IJ}) is simply sinusoidal, for any value of the field angle $\theta$:
\begin{equation}
\label{IJ1}
I_J^S(H,\theta,\phi_0)=J_c A_s \mathcal{F}_S(H, \theta) \sin \phi_0,
\end{equation}

\noindent where we defined:

\begin{equation}
\label{F}
\mathcal{F}_S(H, \theta)\equiv \frac{1}{A_s}\int_S \cos\left[\kappa H(x\cos\theta-y\sin\theta)\right]\, dS.
\end{equation}

\noindent If we decompose the \jun surface, $S$, in two non-overlapping surfaces, $s$ and $s'$, then
the total current, $I_J^S$, is given by the sum of the currents in each surface:

\begin{equation}
\label{IJ2}
I_J^S(H,\theta,\phi_0)=I_J^s(H,\theta,\phi_0)+I_J^{s'}(H,\theta,\phi_0).
\end{equation}  

\noindent If and only if $s$ and $s'$ are point-symmetric concentric figures, $I_J^{s'}$ and $I_J^{s'}$ have the same dependence on $\phi_0$ as $I_J^S$ in Eq.(\ref{IJ1}), Therefore, Eq.(\ref{IJ2})	can be rewritten as:

\begin{equation}
\label{IJ3}
I_J^S(H,\theta,\phi_0)=J_c \left[A_s \mathcal{F}_s(H, \theta)+ A_{s'} \mathcal{F}_{s'}(H, \theta)\right] \sin \phi_0,
\end{equation}

\noindent with $\mathcal{F}_s$ and $\mathcal{F}_{s'}$ defined as in Eq.(\ref{F}). We observe here that upon a rotation around the center of symmetry of an angle $\gamma$ relative to the $Y$-axis, $\mathcal{F}(H, \theta)$ transforms to $\mathcal{F}(H,\theta-\gamma)$, i.e., rotations do not affect the $\phi_0$-dependence of $I_J^S$. Inserting Eq.(\ref{IJ3}) into Eq.(\ref{Ic}), we readily get Eq.(\ref{sum}). Eq.(\ref{diff}) can be derived in a similar fashion, if a surface $s'$ is the relative complement of $s$ in $S$, with $s$ and $S$ point-symmetric and concentric plane figures.

%
%


%
%

\vskip 6pt

As an application of the above theory, let us consider a JTJ shaped in a rectangular annulus, as that shown in Figure~\ref{rectannulus}, upon the assumption that its widths are much smaller than its mean dimensions $\bar{w}$ and $\bar{l}$, i.e., $2\Delta w \equiv W-w << (W+w)/2 \equiv \bar{w}$ and $2 \Delta l=L-l << (L+l)/2 \equiv \bar{l}$. Under such small widths approximation, we can readily compute the MDP of the annular junction, inserting Eq.(\ref{FR}) in Eq.(\ref{diff}): $I_c^{\Delta R}(H,\theta)=J_c A_{\Delta R}\mid \mathcal{F}_{\Delta R}(H,\theta)\mid$ where $A_{\Delta R}=2(\bar{w} \Delta l+\bar{l} \Delta w)$ is the annulus area and 

$$\mathcal{F}_{\Delta R}(H,\theta)= \frac{\Delta l \cos[(\kappa H \bar{l}\cos\theta)/2] \sin[(\kappa H \bar{w}\sin\theta)/2]}{[\kappa H (\bar{w} \Delta l+\bar{l} \Delta w) \sin\theta]/2}+$$
$$+\frac{\Delta w \cos[(\kappa H \bar{w}\sin\theta)/2]\sin[(\kappa H \bar{l}\cos\theta)/2]}{[\kappa H (\bar{w} \Delta l+\bar{l} \Delta w)\cos\theta]/2}.$$

\noindent If the field is applied along the annulus diagonal, then $\bar{w}\sin\bar{\theta}= \bar{l}\cos\bar{\theta}=\bar{p}_R/2 \equiv p_R(\bar{\theta})/2$ and the last expression, independently of $\Delta l$ and $\Delta w$, reduces to a (single) Fraunhofer pattern:

$$\bar{\mathcal{F}}_{\Delta R}(H)=\frac{\sin\kappa H \bar{p}_R/2}{\kappa H \bar{p}_R/2}.$$

\noindent $p_R(\theta)=\bar{w}\sin{\theta}+ \bar{l}\cos{\theta}$ is the length of the projection of the rectangle with sides $\bar{w}$ and $\bar{l}$ in the direction normal to the magnetic field. For a rectangular junction with the field along the diagonals the characteristic function can be derived from Eq.(\ref{FR}):

$$\bar{\mathcal{F}}_{R}(H)=\left( \frac{\sin\kappa H \bar{p}_R/4}{\kappa H \bar{p}_R/4} \right)^2.$$ 

\noindent Therefore, the MDP of a narrow rectangular annular JTJ, when compared to that of a full rectangular junction, although qualitatively similar, has an halved periodicity and a slower sidelobes suppression. 

\vskip 6pt

If $s$ and $s'$ are not concentric, Eqs.(\ref{diff}) and (\ref{sum}) only apply when the magnetic field is aligned with the two centers of symmetry, that is, when the field angle is $\bar{\theta}=\arctan x_c/y_c$, where $(x_c,y_c)$ are the coordinates of one center relative to the other. In fact, upon a translation $(x,y)=(x'+x_c,y'+y_c)$, the quantity $(x\cos\theta- y\sin\theta)$ in Eq.(\ref{F})	transforms to $(x'\cos\theta-y'\sin\theta+x_c\cos\theta -y_c\sin\theta)$ and $x_c\cos\theta-y_c\sin\theta=0$ only when $\tan\bar{\theta}=x_c/y_c$. In other words, the translation of the inner boundary of an annular junction along the direction of the applied field does not alter the its MDP. A similar reasoning can be applied to multiply connected \Jos tunnel junctions whose surface is determined by an outer boundary and several inner boundary of point symmetric plane figures with centers of symmetry aligned with the magnetic field.

\vskip 6pt
The symmetry properties were not envisages in previous works dealing with point symmetric JTJs and in which the calculated MDPs were not cast in the form given in Eq.(\ref{diff}) \cite{NappiPRB97}. Furthermore, the additive properties can also be invoked to simplify the computation of the magnetic dependence of the amplitudes of the Fiske-resonances which also involves surface integrals.


\begin{thebibliography}{99}
\bibitem{note1} It can be be demonstrated that, if the rectangle is rotated by an arbitrary angle $\gamma$ with respect to the $Y$-axis, as intuitively expected, Eq.(\ref{IcRectangle}) transforms to:$$I_c^R(H,\theta)=J_c A_R \left| \frac{\sin\left\{[\kappa H W \sin(\theta-\gamma)]/2\right\}} {[\kappa H W \sin(\theta-\gamma)]/2} \times \right.$$
 $$\left. \times \frac{\sin\left\{[\kappa H L \cos(\theta-\gamma)]/2\right\}} {[\kappa H L \cos(\theta-\gamma)]/2} \right|.$$
\end{thebibliography}
\end{document}